# EQNN: 改進式量子神經網路

Abel C. H. Chen
*Information & Communications Security Laboratory,*
*Chunghwa Telecom Laboratories*
Taoyuan, Taiwan
ORCID: 0000-0003-3628-3033; Email: chchen.scholar@gmail.com

*摘要*—隨著量子計算技術的成熟，許多研究也逐漸轉向探索量子計算的應用。搭配人工智慧的熱潮，各種機器學習方法也陸續被開發成量子電路和量子演算法。其中，量子神經網路可以通過特徵圖(Feature Map)把輸入映射到量子電路，並且通過變分模型(Variational Model)調整參數值，可以應用在迴歸和分類等。然而，如何設計一個適合應用問題的特徵圖是重要的挑戰之一。有鑑於此，本研究提出一個改進式量子神經網路(Enhanced Quantum Neural Network, EQNN)，並且該改進式量子神經網路包含本研究設計的改進式特徵圖(Enhanced Feature Map, EFM)，可以有效把輸入變量映射到更適合量子計算的值域，作為變分模型的輸入來提升正確率。在實驗環境中，本研究以行動資料用量預測為例，可以根據使用者的行動資料用量來推薦適合的費率方案。本研究提出的改進式量子神經網路與現行主流的量子神經網路進行對比，實驗結果顯示改進式量子神經網路可以用更少的量子邏輯閘達到更高的正確率，並且在不同的最佳化演算法下都能更快收斂到最佳解。

*關鍵字*—改進式量子神經網路、特徵圖、變分模型、行動資料用量、實證分析

## I. 前言

近年來在量子計算技術的不斷進步下，許多研究也陸續在量子特性(包含量子疊加態和量子糾纏態)[1]的基礎上來探索指數級加速計算的可能性，嘗試把量子計算用來解決一些應用場域的問題。其中，量子機器學習[2]是目前發展的主流應用之一，包含有量子 k 個最近鄰居演算法[3]、量子貝氏網路[4]、量子支持向量機[5]、量子神經網路[6]等。特別是近幾年在神經網路和深度學習的發展熱潮下，越來越多人重視量子神經網路的發展潛力，並且也開發出量子卷積神經網路[7]、量子循環神經網路[8]、量子生成對抗網路[9]等模型，許多研究都指向量子神經網路逐漸成為研究熱點，並且未來有機會發展出殺手級應用。

雖然目前已經有發展出量子神經網路相關套件(如：IBM Qiskit)和各種量子最佳化演算法(如：線性近似約束最佳化(Constrained Optimization By Linear Approximation, COBYLA)[10]、同時擾動隨機近似(Simultaneous Perturbation Stochastic Approximation, SPSA)[11]、解析量子梯度下降(Analytic Quantum Gradient Descent, AQGD)[12]等)，可供大家在此基礎上開發。然而，量子神經網路的結構主要包含特徵圖(Feature Map)和變分模型(Variational Model)，如何設計合適的特徵圖可以把應用場域的輸入變量映射到合適的量子電路，以及如何設計合適的變分模型可以讓量子神經網路在擬合應用場域預測值時可以得到更低的損失值，這些都是量子神經網路的主要挑戰。

有鑑於此，本研究將比較量子神經網路與傳統神經網路激活函數，分析特徵圖映射的值域變化，從而設計改進式特徵圖(Enhanced Feature Map, EFM)。除此之外，本研究提出改進式量子神經網路(Enhanced Quantum Neural Network, EQNN)結合改進式特徵圖和變分模型，並且應用在行動資料用量預測。為證明本研究提出的改進式量子神經網路之效率和效能，主要對比現行主流的量子神經網路，並且採用不同的量子最佳化演算法，觀察實驗結果的變化。綜合上述，本研究的主要貢獻條列如下：

- 本研究分析量子神經網路擬合傳統神經網路激活函數的效果，並從而設計出改進式特徵圖。
- 本研究在提出的改進式特徵圖基礎上，結合變分模型，設計改進式量子神經網路。
- 本研究採用不同的量子最佳化演算法來比較改進式量子神經網路和最現代的(State-of-the-Art, SOTA)方法，並證實本研究提出的改進式量子神經網路可以用更少的量子邏輯閘來達到更高的正確率。

本論文主要分為六個小節。第 II 節將介紹量子神經網路的結構，並採用 ZZFeatureMap 特徵圖為例[6]。第 III 節比較量子神經網路與傳統神經網路常見的激活函數，並證明量子神經網路也能達到同等效果。第 IV 節介紹本研究提出的改進式特徵圖和改進式量子神經網路的具體量子電路和量子演算法。第 V 節描述實驗環境和討論比較結果。最後，第 VI 節整理本研究貢獻和討論未來方向。

## II. 量子神經網路

本節將先介紹量子神經網路的整體架構，再描述量子態定義及量子神經網路中會用到的邏輯閘。最後再各別展開特徵圖的結構和變分模型的結構來介紹。

### A. 整體架構

量子神經網路的架構主要包含輸入、特徵圖、變分模型、以及測量，如圖 1 所示。

假設輸入總共包含 $n$ 個量子位元(即$|q_0\rangle$、$|q_1\rangle$、…、$|q_{n-1}\rangle$)，量子態表示為$|q_{n-1} \cdots q_1 q_0\rangle$。在量子神經網路輸入量子態的初始狀態與輸入變量$X$(該輸入變量$X = \{x_0, x_1, …, x_{n-1}\}$)無關。

特徵圖是由特徵圖運算子$\alpha(X)$重覆 $r_1$ 次(即$\alpha(X)^{r_1}$)所構成，主要用來把輸入變量映射到量子電路。因此，輸入變量 $X$ 作為特徵圖運算子$\alpha(X)$的參數來調整量子態，把輸入變量映射到量子電路。特徵圖運算子的詳細量子電路結構將在第 II.C 節說明。

變分模型是由變分模型運算子$\beta(W)$重覆 $r_2$ 次(即$\beta(W)^{r_2}$)所構成，主要建立權重值用以產生預測值。因此，權重變量 $W$ (該權重變量 $W = \{w_0, w_1, …, w_{m-1}\}$)作為變分模型運算子$\beta(W)$的參數來調整量子態，根據經過特徵圖計算後的量子態進行調整產生預測值。變分模型運算子的詳細量子電路結構將在第 II.D 節說明。

學習過程中可以採用不同的量子最佳化演算法調整變分模型中的權重變量值，根據定義的損失函數(或稱為目標函數)，讓產生出來的預測值盡可能接近真值，降低損失值。

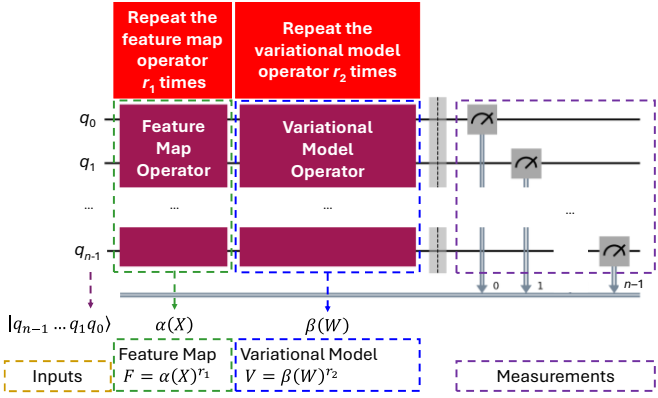

Fig. 1. 量子神經網路架構圖

*B. 量子態和常用邏輯閘*

本節定義單量子位元$|q_i\rangle$的量子態，雙量子位元$|q_jq_i\rangle$的量子態，以及常用邏輯閘包含 Hadamard Gate、Phase Gate、RY Gate、Cnot Gate，其數學表達式和符號如表 I 所示。

TABLE I. QUANTUM LOGIC GATES FOR A QNN

| Logic Gate | Notation | Symbol |
| --- | --- | --- |
| Hadamard Gate | $H\|q_i\rangle = \frac{1}{\sqrt{2}}\begin{bmatrix}1 & 1 \\ 1 & -1\end{bmatrix}\|q_i\rangle$ | $q_i$ —[H]— |
| Phase Gate | $P(\theta)\|q_i\rangle = \begin{bmatrix}1 & 0 \\ 0 & e^{i\theta}\end{bmatrix}\|q_i\rangle$ | $q_i$ —[P$_\theta$]— |
| RY Gate | $R(\theta)\|q_i\rangle = \begin{bmatrix}cos\left(\frac{\theta}{2}\right) & -sin\left(\frac{\theta}{2}\right) \\ sin\left(\frac{\theta}{2}\right) & cos\left(\frac{\theta}{2}\right)\end{bmatrix}\|q_i\rangle$ | $q_i$ —[R$_{Y,\theta}$]— |
| Cnot Gate Control Qubit: $q_i$ Target Qubit: $q_j$ | $C(q_i)\|q_jq_i\rangle = \begin{bmatrix}1 & 0 & 0 & 0 \\ 0 & 0 & 0 & 1 \\ 0 & 0 & 1 & 0 \\ 0 & 1 & 0 & 0\end{bmatrix}\|q_jq_i\rangle$ | $q_i$ —•— $q_j$ —⊕— |

單量子位元$|q_i\rangle$量子態的表達方式如公式(1)所示，由$\begin{bmatrix}v_0 \\ v_1\end{bmatrix}$表示為量子位元$|q_i\rangle$的振幅向量，並且可由$v_0^2$表示為發生$|0\rangle$的機率、由$v_1^2$表示為發生$|1\rangle$的機率。因此，當$|q_i\rangle = |0\rangle$，則振幅向量為$\begin{bmatrix}1 \\ 0\end{bmatrix}$，如公式(2)所示。同理，當$|q_i\rangle = |1\rangle$，則振幅向量為$\begin{bmatrix}0 \\ 1\end{bmatrix}$，如公式(3)所示[13]-[14]。

$$|q_i\rangle = v_0|0\rangle + v_1|1\rangle \rightarrow \begin{bmatrix}v_0 \\ v_1\end{bmatrix} \quad (1)$$

$$|q_i\rangle = |0\rangle = 1|0\rangle + 0|1\rangle \rightarrow \begin{bmatrix}1 \\ 0\end{bmatrix} \quad (2)$$

$$|q_i\rangle = |1\rangle = 0|0\rangle + 1|1\rangle \rightarrow \begin{bmatrix}0 \\ 1\end{bmatrix} \quad (3)$$

當量子位元$|q_i\rangle$通過 Hadamard Gate 後可以產生均勻疊加態，如公式(4)所示。在 Phase Gate，可以對布洛赫球(Bloch sphere)的 Z 軸旋轉角度$\theta$來產生不同的相位差，如公式(5)所示。RY Gate 則是通過對布洛赫球的 Y 軸旋轉角度$\theta$來產生不同的相位差，如公式(6)所示[13]-[14]。

$$H|q_i\rangle = \frac{1}{\sqrt{2}}\begin{bmatrix}1 & 1 \\ 1 & -1\end{bmatrix}\begin{bmatrix}v_0 \\ v_1\end{bmatrix} = \frac{1}{\sqrt{2}}\begin{bmatrix}v_0 + v_1 \\ v_0 - v_1\end{bmatrix} \quad (4)$$

$$P(\theta)|q_i\rangle = \begin{bmatrix}1 & 0 \\ 0 & e^{i\theta}\end{bmatrix}\begin{bmatrix}v_0 \\ v_1\end{bmatrix} = \begin{bmatrix}v_0 \\ v_1 e^{i\theta}\end{bmatrix} \quad (5)$$

$$\begin{aligned}R(\theta)|q_i\rangle &= \begin{bmatrix}cos\left(\frac{\theta}{2}\right) & -sin\left(\frac{\theta}{2}\right) \\ sin\left(\frac{\theta}{2}\right) & cos\left(\frac{\theta}{2}\right)\end{bmatrix}\begin{bmatrix}v_0 \\ v_1\end{bmatrix} \\ &= \begin{bmatrix}v_0 cos\left(\frac{\theta}{2}\right) - v_1 sin\left(\frac{\theta}{2}\right) \\ v_0 sin\left(\frac{\theta}{2}\right) + v_1 cos\left(\frac{\theta}{2}\right)\end{bmatrix}\end{aligned} \quad (6)$$

雙量子位元$|q_jq_i\rangle$量子態的表達方式如公式(7)所示，由$\begin{bmatrix}v_{00} \\ v_{01} \\ v_{10} \\ v_{11}\end{bmatrix}$表示為量子位元$|q_jq_i\rangle$的振幅向量，並且可由$v_{00}^2$表示為發生$|00\rangle$的機率，如公式(8)所示[13]-[14]。同理，可得$|01\rangle$、$|10\rangle$、$|11\rangle$量子態的振幅向量，由於篇幅限制，不展開描述。

$$\begin{aligned}|q_jq_i\rangle &= v_{00}|00\rangle + v_{01}|01\rangle + v_{10}|10\rangle + v_{11}|11\rangle \\ &\rightarrow \begin{bmatrix}v_{00} \\ v_{01} \\ v_{10} \\ v_{11}\end{bmatrix}\end{aligned} \quad (7)$$

$$|00\rangle = 1|00\rangle + 0|01\rangle + 0|10\rangle + 0|11\rangle \rightarrow \begin{bmatrix}1 \\ 0 \\ 0 \\ 0\end{bmatrix} \quad (8)$$

當量子位元$|q_jq_i\rangle$通過 Cnot Gate 時，並且以量子位元$|q_i\rangle$為控制位元，且量子位元$|q_j\rangle$為目標位元，則會根據量子位元$|q_i\rangle$為$|1\rangle$時來調整量子態，如公式(9)所示[13]-[14]。可以觀察到振幅向量$v_{01}$和$v_{11}$互換。

$$C(q_i)|q_jq_i\rangle = \begin{bmatrix}1 & 0 & 0 & 0 \\ 0 & 0 & 0 & 1 \\ 0 & 0 & 1 & 0 \\ 0 & 1 & 0 & 0\end{bmatrix}\begin{bmatrix}v_{00} \\ v_{01} \\ v_{10} \\ v_{11}\end{bmatrix} = \begin{bmatrix}v_{00} \\ v_{11} \\ v_{10} \\ v_{01}\end{bmatrix} \quad (9)$$

*C. 特徵圖*

目前已經有發展出多個不同的特徵圖運算子，其中常見的特徵圖運算子包含有 PauliFeatureMap、ZFeatureMap、ZZFeatureMap [6], [15]等。本節為說明特徵圖的量子電路結構，主要參考 IBM 提供的量子神經網路實例[16]，採用 ZZFeatureMap 作為特徵圖運算子$\alpha(X)$，並且根據實例的作法該特徵圖運算子僅重覆 1 次(即 $r_1 = 1$)，故特徵圖為 $F = \alpha(X)$。並且在實驗時，將以此結構作為基準(benchmark)來衡量本研究提出的改進式特徵圖。

在兩個量子位元時，特徵圖的量子電路結構如圖 2 所示。首先將先對每個量子位元進行 Hadamard Gate 操作以產生各種量子態的均勻疊加態，再對每個量子位元進行 Phase Gate 操作以產生不同的相位差，後續運用 Cnot Gate 來該量子位元之間可以交互影響，產生量子糾纏態。在此例中，總共用到 7 個量子邏輯閘來計算。其中，在此例中有兩個輸入變量 $X = \{x_0, x_1\}$，並且輸入變量 $x_0$ 作為作用在量子位元$|q_0\rangle$的 Phase Gate 的角度參數，而輸入變量 $x_1$ 作為作用在量子位元$|q_1\rangle$的 Phase Gate 的角度參數，以及同時採用輸入變量 $x_0$ 和 $x_1$ 作為在兩個 Cnot Gate 之間的 Phase Gate 的角度參數。

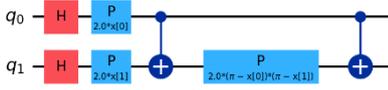

Fig. 2. 特徵圖

### D. 變分模型

目前已經有發展出多個不同的變分模型運算子，其中常見的變分模型運算子包含有 RealAmplitudes、FidelityQuantumKernel [6], [15]等。本節為說明變分模型的量子電路結構，主要參考 IBM 提供的量子神經網路實例[16]，採用 RealAmplitudes 作為變分模型運算子$\beta(W)$，並且根據實例的作法該特徵圖運算子重覆 3 次(即 $r_2 = 3$)，故特徵圖為$V = \beta(W)^3$。並且在實驗時，將以此結構作為基準(benchmark)來建立量子神經網路。

在兩個量子位元時，首先將先對每個量子位元進行 RY Gate 操作以產生不同的相位差，再運用 Cnot Gate 來該量子位元之間可以交互影響，產生量子糾纏態，如圖 3 為 RealAmplitudes 迭代一次的量子電路結構。當迭代 3 次後的量子電路結構如圖 4 所示，也是本研究的基準(benchmark)變分模型。在此例中，總共用到 11 個量子邏輯閘來計算。其中，在此例中有 8 個權重變量 $W = \{w_0, w_1, …, w_8\}$，並且權重變量 $w_0$、$w_2$、$w_4$、$w_6$ 作為作用在量子位元$|q_0\rangle$的 Phase Gate 的角度參數，而權重變量$w_1$、$w_3$、$w_5$、$w_7$作為作用在量子位元$|q_1\rangle$的 Phase Gate 的角度參數，每個 Phase Gate 操作之間採用 Cnot Gate 操作以產生量子位元之間可以交互影響。

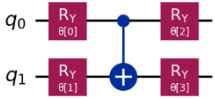

Fig. 3. 變分模型(迭代 1 次 RealAmplitudes)

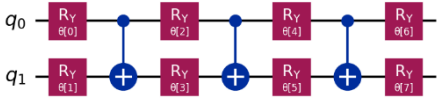

Fig. 4. 變分模型(迭代 3 次 RealAmplitudes)

## III. 量子神經網路邏輯閘對比傳統神經網路激活函數

本節主要比較量子神經網路邏輯閘和傳統神經網路激活函數。首先先提出簡化版量子神經網路，並從理論說明運作原理，再證明此簡化版量子神經網路可以達到與傳統神經網路激活函數同等效果。

### A. 簡化版量子神經網路模型

本節中提出的簡化版量子神經網路只有 1 個量子位元，並且特徵圖只有 1 個 Hadamard Gate 和 1 個 RY Gate (即對應的輸入變量 $x$ 只有 1 個)、變分模型只有 1 個 RY Gate (即對應的權重變量 $w$ 只有 1 個)，簡化版量子神經網路的量子電路如圖 5 所示。其中，假設輸入的初始量子態 $|q\rangle = |0\rangle$(即$\begin{bmatrix}v_0\\v_1\end{bmatrix}=\begin{bmatrix}1\\0\end{bmatrix}$)，當經過特徵圖計算後可得公式(10)，再經過變分模型計算後可得公式(11)。當在迴歸應用時，可以運用公式(12)產生預測值，並以公式(13)計算該筆資料的損失值(假設平方誤差為損失函數)。

$$
\begin{aligned}
R(x)H|q\rangle &= \begin{bmatrix} \cos\left(\frac{x}{2}\right) & -\sin\left(\frac{x}{2}\right) \\ \sin\left(\frac{x}{2}\right) & \cos\left(\frac{x}{2}\right) \end{bmatrix} \frac{1}{\sqrt{2}}\begin{bmatrix} 1 & 1 \\ 1 & -1 \end{bmatrix}\begin{bmatrix} 1 \\ 0 \end{bmatrix} \\
&= \frac{1}{\sqrt{2}}\begin{bmatrix} \cos\left(\frac{x}{2}\right) & -\sin\left(\frac{x}{2}\right) \\ \sin\left(\frac{x}{2}\right) & \cos\left(\frac{x}{2}\right) \end{bmatrix}\begin{bmatrix} 1 \\ 1 \end{bmatrix} \\
&= \frac{1}{\sqrt{2}}\begin{bmatrix} \cos\left(\frac{x}{2}\right) - \sin\left(\frac{x}{2}\right) \\ \sin\left(\frac{x}{2}\right) + \cos\left(\frac{x}{2}\right) \end{bmatrix}
\end{aligned} \tag{10}
$$

$$y' = \rho_0^2 - \rho_1^2 \tag{12}$$
$$L_{regression} = (y' - y)^2 \tag{13}$$

當在分類應用時，$\rho_0^2$可以運用作為 Label 0 的機率預測值，以及$\rho_1^2$運用作為 Label 1 的機率預測值，並以公式(14)計算該筆資料的損失值(假設交叉熵為損失函數，Label 0 的機率真值為$p_0$、Label 1 的機率真值為$p_1$)。

$$L_{classification} = -(p_0 \log \rho_0^2 + p_1 \log \rho_1^2) \tag{14}$$

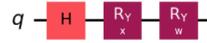

Fig. 5. 簡化版量子神經網路

### B. 擬合線性函數

線性函數定義如公式(15)所示，本節的資料集採用介於[–1, 1]區間的 200 筆隨機數作為 $x$ 值，並且代入公式(15)得到對應的 $y$ 值。運用簡化版量子神經網路學習此資料集，訓練後的 $w$ 值為 3.14150444，擬合結果如圖 6 所示。可以觀察到大部分的資料都以擬合，並有較小的誤差。並且可從公式驗證，例如：$w = 3.14150444$ 且 $x = 0.1$ 代入公式(11)和(12)可得預測 $y$' 為 0.099833417，近似真值 $y = 0.1$。

$$y = l(x) = x \tag{15}$$

$$
\begin{aligned}
R(w)R(x)H|q\rangle &= \begin{bmatrix} \cos\left(\frac{w}{2}\right) & -\sin\left(\frac{w}{2}\right) \\ \sin\left(\frac{w}{2}\right) & \cos\left(\frac{w}{2}\right) \end{bmatrix} \frac{1}{\sqrt{2}}\begin{bmatrix} \cos\left(\frac{x}{2}\right) - \sin\left(\frac{x}{2}\right) \\ \sin\left(\frac{x}{2}\right) + \cos\left(\frac{x}{2}\right) \end{bmatrix} \\
&= \frac{1}{\sqrt{2}}\begin{bmatrix} \cos\left(\frac{w}{2}\right)\left(\cos\left(\frac{x}{2}\right) - \sin\left(\frac{x}{2}\right)\right) - \sin\left(\frac{w}{2}\right)\left(\sin\left(\frac{x}{2}\right) + \cos\left(\frac{x}{2}\right)\right) \\ \sin\left(\frac{w}{2}\right)\left(\cos\left(\frac{x}{2}\right) - \sin\left(\frac{x}{2}\right)\right) + \cos\left(\frac{w}{2}\right)\left(\sin\left(\frac{x}{2}\right) + \cos\left(\frac{x}{2}\right)\right) \end{bmatrix} = \begin{bmatrix} \rho_0 \\ \rho_1 \end{bmatrix}
\end{aligned} \tag{11}
$$

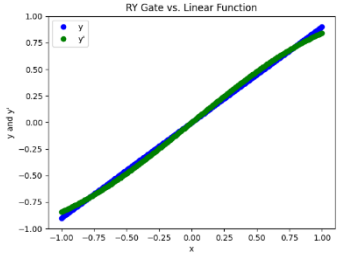

Fig. 6. 線性函數擬合結果

## C. 擬合 Sigmoid 函數

Sigmoid 函數定義如公式(16)所示，本節的資料集採用介於[–3, 3]區間的 200 筆隨機數作為 x 值，並且代入公式(16)和(17)得到對應的 y 值。其中，由於量子神經網路迴歸應用的預測值的值域為[–1, 1](參考公式(12))，所以運用公式(17)修改 y 值的值域為[–1, 1]。除此之外，把 x 同除以 2，讓輸入到簡化版量子神經網路的輸入變量值域介於[–1.5, 1.5]，擬合結果如圖 7 所示。可以觀察到有較好的擬合結果，僅有很小的誤差。

$$s(x) = \frac{1}{1+e^{-x}} \tag{16}$$
$$y = 2 \times s(x) - 1 \tag{17}$$

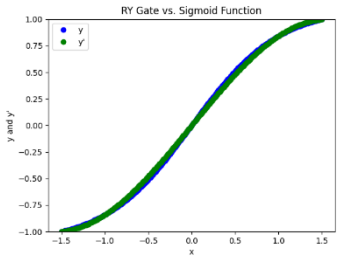

Fig. 7. Sigmoid 函數擬合結果

## D. 擬合 Tanh 函數

Tanh 函數定義如公式(18)所示，本節的資料集採用介於[–1.5, 1.5]區間的 200 筆隨機數作為 x 值，並且代入公式(18)得到對應的 y 值。運用簡化版量子神經網路學習此資料集，擬合結果如圖 8 所示。由此結果可知，量子邏輯閘對 Tanh 函數也可以有很好的擬合結果，證明量子神經網路也有能做到和傳統神經網路相同的效果。

$$y = t(x) = \frac{e^x - e^{-x}}{e^x + e^{-x}} \tag{18}$$

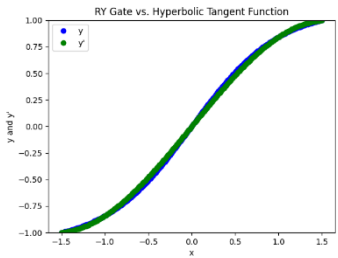

Fig. 8. Tanh 函數擬合結果

## IV. 本研究提出的改進式量子神經網路

根據第 III 節比較量子神經網路邏輯閘和傳統神經網路激活函數的結果，本研究提出改進式特徵圖，並且在改進式特徵圖基礎上結合變分模型建立改進式量子神經網路。由於本研究採用行動資料用量預測案例，主要採用使用者同一個月的兩次行動資料用量(月中統計一次和月底統計一次)作為輸入變量(即兩個輸入變量)，然後預測使用者適合哪一種方案(即分類應用)，所以以下說明皆採用兩個量子位元為例來建構改進式量子神經網路。

### A. 本研究提出的改進式特徵圖

在傳統神經網路中，激活函數 Sigmoid 函數和 Tanh 函數都在分類應用中起到了卓越的表現。有鑑於量子神經網路擬合 Sigmoid 函數和擬合 Tanh 函數的結果發現，x 的值域介於[–1.5, 1.5]區間會更合適。因此，本研究在資料預處理先對 x 做正規化為[0, 1]區間，再作為輸入變量輸入到 RY Gate，並且以公式(19)調整值域為[–1.5, 1.5]區間。

$$2 \times x - 1.5 \tag{19}$$

本研究提出的改進式特徵圖的量子電路結構如圖 9 所示，將對輸入的量子位元先各別操作 Hadamard Gate 產生量子疊加態，之後再對每個量子位元依上述的方式操作 RY Gate，並把修改後的 x 值作為輸入變量(即 GY Gate 的角度值)，最後再操作 Cnot Gate 建立量子糾纏態。本研究提出的改進式特徵圖的量子態變化數學式可參考公式(20)。總共僅用到 5 個量子邏輯閘。

$$C(q_0)\big(R(2x_0 - 1.5) \otimes R(2x_1 - 1.5)\big)(H \otimes H)|q_1 q_0\rangle \tag{20}$$

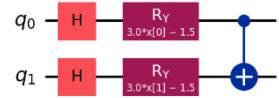

Fig. 9. 本研究提出的改進式特徵圖

### B. 本研究提出的 EQNN 模型 1: *EFM + 1 次 RealAmplitudes*

本研究提出的改進式量子神經網路主要結合本研究提出的改進式特徵圖和現行主流的變分模型運算子。在改進式量子神經網路模型 1 中，本研究採用 RealAmplitudes 作為變分模型運算子，並且只迭代 1 次(即 $r_2 = 1$)，依此作為變分模型。改進式量子神經網路模型 1 的量子電路結構如圖 10 所示，可以觀察到變分模型用到 5 個量子邏輯閘，加上改進式特徵圖的 5 個量子邏輯閘，合計共用到 10 個量子邏輯閘。

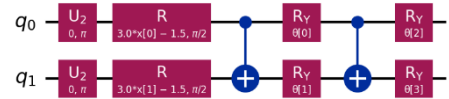

Fig. 10. 改進式量子神經網路模型 1

### C. 本研究提出的 EQNN 模型 2 和模型 3:

有鑑於 IBM 提供的量子神經網路實例[6], [16](即基準(benchmark))在變分模型的結構中採用了 3 次 RealAmplitudes，所以為了公平且完整比較，本研究考慮一併比較 2 次 RealAmplitudes 和 3 次 RealAmplitudes。

本研究提出的改進式量子神經網路模型 2 主要採用改進式特徵圖作為特徵圖的結構，並且在變分模型結構主

要由迭代 2 次(即 $r_2 = 2$)的 RealAmplitudes 所組合,如圖 11 所示。在模型 2 中包含了 13 個量子邏輯閘(即 13 = 5 + 8)。

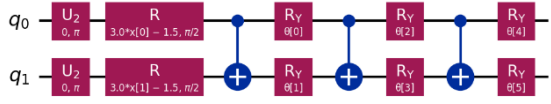

Fig. 11. 改進式量子神經網路模型 2

本研究提出的改進式量子神經網路模型 3 主要採用改進式特徵圖結合迭代 3 次 RealAmplitudes 的變分模型(如圖 4)所組成的量子電路,如圖 12 所示。在模型 3 中包含了 16 個量子邏輯閘(即 16 = 5 + 11)。

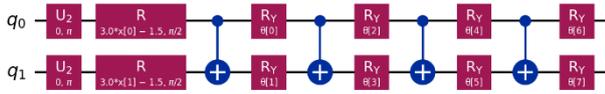

Fig. 12. 改進式量子神經網路模型 3

## V. 實驗環境與實證結果

為驗證本研究提出的改進式特徵圖和改進式量子神經網路,本研究採用以行動資料用量預測為例來分析,並將先介紹資料集和實驗環境。在實驗結果中比較每個模型的量子邏輯閘數量、分類正確率、以及收斂速度,從而驗證效率和效能不同面向。

### A. 實驗環境

本研究採用的行動資料用量主要來自[17]和[18],在每個月月中統計每個使用者的行動資料用量得到 $x_0$ 和在每個月月底統計每個使用者的行動資料用量得到 $x_1$,作為輸入變量,並且把該使用者採用的方案作為輸出(即分類類別 $y_0$ 和 $y_1$)。在本研究中,資料集主要收集類別 $y_0$ 資料 500 筆和 $y_1$ 資料 500 筆。

為實作各種量子神經網路模型,本研究採用 IBM Qiskit 套件開發,並且量子最佳化演算法上主要採用線性近似約束最佳化(COBYLA)、同時擾動隨機近似(SPSA)、解析量子梯度下降(AQGD)進行比較。在每個模型的訓練過程中,採用的最佳化演算法都各別迭代 100 次。

### B. 實證結果與討論

本節採用 IBM 提供的量子神經網路實例[6], [16]作為基準(Benchmark),並與本研究提出的 EQNN 模型 1、EQNN 模型 2、EQNN 模型 3 進行比較。

首先比較每個模型的量子邏輯閘數量,比較結果如表 II 所示。由於基準模型的特徵圖採用 ZZFeatureMap 運算子,用了 7 個量子邏輯閘,加上迭代 3 次 RealAmplitudes 的變分模型有 11 個量子邏輯閘,所以總共用了 18 個量子邏輯閘。本研究提出的 EQNN 模型 1 採用改進式特徵圖,只用 5 個量子邏輯閘,並且可以僅搭配迭代 1 次 RealAmplitudes 的變分模型有 5 個量子邏輯閘,所以總共只用了 10 個量子邏輯閘。同理,本研究提出的 EQNN 模型 2 和 EQNN 模型 3 分別只用了 13 個、16 個量子邏輯閘。因此,本研究提出的 3 個模型使用到的量子邏輯閘數量都較基準模型型使用到的量子邏輯閘數量少,所以計算速度更快。

TABLE II. 每個模型的量子邏輯閘數量

|  | Benchmark [6], [16] | EQNN Model 1 | EQNN Model 1 | EQNN Model 3 |
|---|---|---|---|---|
| Feature Map | 7 | 5 | 5 | 5 |
| Variational Model | 11 | 5 | 8 | 11 |
| Summary | 18 | 10 | 13 | 16 |

在本研究中,為表 III 整理每個模型的分類正確率,由實驗結果可以觀察到,不論採用哪一種量子最佳化演算法,本研究提出的 3 個模型都能達到 100%的正確率。然而,基準模型卻無法達到 100%的正確率;當採用 COBYLA 時,正確率達到 97.9%、當採用 SPSA 時,正確率達到 99.3%、當採用 AQGD 時,正確率達到 99.9%。

TABLE III. 每個模型的分類正確率

| Optimization Algorithm | Benchmark [6], [16] | EQNN Model 1 | EQNN Model 1 | EQNN Model 3 |
|---|---|---|---|---|
| COBYLA | 97.9% | 100.0% | 100.0% | 100.0% |
| SPSA | 99.3% | 100.0% | 100.0% | 100.0% |
| AQGD | 99.9% | 100.0% | 100.0% | 100.0% |

為公平比較,本研究以 AQGD 量子最佳化演算法為例,觀察不同模型的收斂速度,由圖 13 可以看到基準模型在第 20 次迭代時損失值降低速度趨緩,而圖 14 可以看到本研究提出的 EQNN 模型 1 在第 60 次迭代時損失值降低速度才趨緩,也就是基準模型的收斂速度比本研究提出的 EQNN 模型 1 快速,但 EQNN 模型 1 收斂結果較好。與 EQNN 模型 1 類似,圖 15 為 EQNN 模型 2 的收斂結果。然而,如果同樣是採用迭代 3 次 RealAmplitudes 的變分模型的 EQNN 模型 3 則表現出卓越的收斂速度,在第 5 次迭代時損失值降低速度就已趨緩,如圖 16 所示。

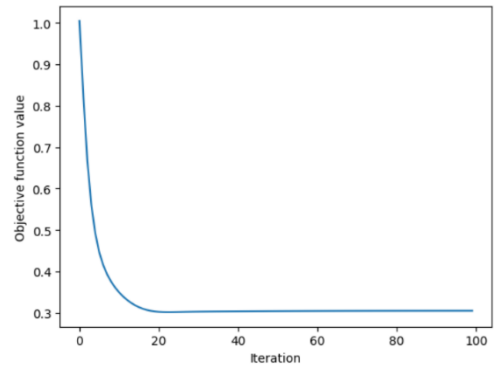

Fig. 13. Benchmark 模型收斂速度

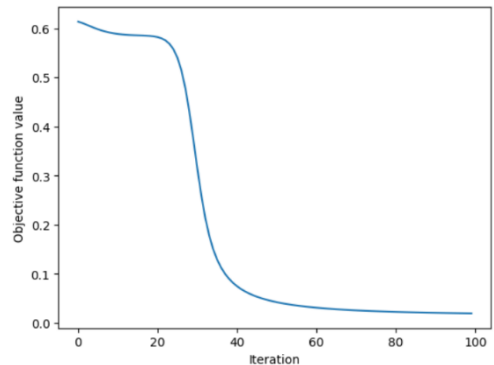

Fig. 14. EQNN 模型 1 收斂速度

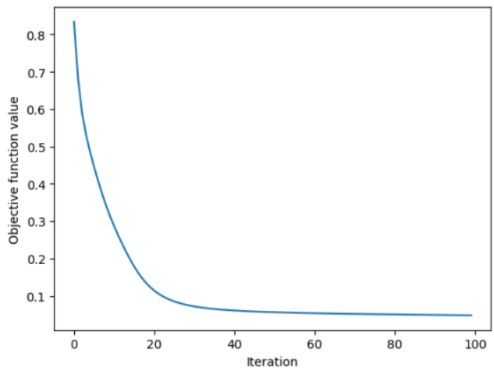

Fig. 15. EQNN 模型 2 收斂速度

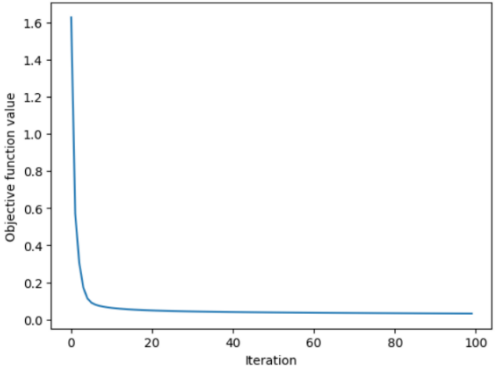

Fig. 16. EQNN 模型 3 收斂速度

## VI. 結論與未來研究

本研究比較量子神經網路量子邏輯閘和傳統神經網路激活函數，並且從而設計了改進式特徵圖。在改進式特徵圖基礎上，本研究提出 3 個改進式較量子神經網路，並且採用真實行動資料用量預測為例，驗證本研究提出的模型可以用更少的量子邏輯閘達到更高的正確率。

未來可嘗試擴展本研究提出的改進式特徵圖支援更多的量子位元，並且應用在不同的領域以驗證改進式特徵圖的通用性。

### 致謝

本研究採用 IBM Qiskit 套件開發，感謝 IBM 免費提供量子計算開發工具。

### 參考文獻